\documentclass[conference]{IEEEtran}
\usepackage{cite}
\usepackage{amsmath,amssymb,amsfonts}
\usepackage{enumitem}
\usepackage{algorithmic}
\usepackage{graphicx}
\usepackage{textcomp}
\usepackage{xcolor}
\def\BibTeX{{\rm B\kern-.05em{\sc i\kern-.025em b}\kern-.08em
    T\kern-.1667em\lower.7ex\hbox{E}\kern-.125emX}}
\usepackage{tikz}
\usetikzlibrary{positioning, shapes.multipart, arrows.meta}

\newcommand{\ttt}[1]{\texttt{#1}}

\usepackage{algorithm2e}
\usepackage[frozencache=true]{minted}
\usepackage{listings}
\begin{document}

\title{Breaking Precision Time: OS Vulnerability Exploits Against IEEE 1588}

\author{\IEEEauthorblockN{Muhammad Abdullah Soomro}
\IEEEauthorblockA{\textit{University of Massachusetts Amherst}\\
msoomro@umass.edu}
\and
\IEEEauthorblockN{Fatima Muhammad Anwar}
\IEEEauthorblockA{\textit{University of Massachusetts Amherst} \\
fanwar@umass.edu}
}

\maketitle

\begin{abstract}
The Precision Time Protocol (PTP), standardized as IEEE 1588, provides sub-microsecond synchronization across distributed systems and underpins critical infrastructure in telecommunications, finance, power systems, and industrial automation. While prior work has extensively analyzed PTP's vulnerability to network-based attacks, prompting the development of cryptographic protections and anomaly detectors, these defenses presume an uncompromised host. In this paper, we identify and exploit a critical blind spot in current threat models: kernel-level adversaries operating from within the host running the PTP stack. We present the first systematic study of kernel-rooted attacks on PTP, demonstrating how privileged attackers can manipulate system time by corrupting key interfaces without altering PTP network traffic. We implement three attack primitives, constant offset, progressive skew, and random jitter, using in-kernel payloads, and evaluate their impact on the widely used \texttt{ptp4l} and \texttt{phc2sys} daemons. Our experiments reveal that these attacks can silently destabilize clock synchronization, bypassing existing PTP security extensions. These findings highlight the urgent need to reconsider host-level trust assumptions and integrate kernel integrity into the design of secure time synchronization systems.
\end{abstract}

\begin{IEEEkeywords}
Precision Time Protocol (PTP), IEEE 1588, time synchronization, kernel security
\end{IEEEkeywords}

\section{Introduction} \label{sec:introduction}

Precision Time Protocol (PTP), defined by the IEEE 1588 standard, has become a cornerstone for time synchronization across diverse industries, ranging from telecommunications \cite{Comcores-PTP-5G-2021} and power grids \cite{FITNESS-PACW-2018} to high-frequency trading \cite{Hoptroff-MiFID-2024} and industrial automation \cite{Rockwell-CIPSync-2023}. By enabling distributed clocks to agree within sub-microsecond or even nanosecond ranges, PTP underpins applications that demand exceedingly tight timing coordination. Over the years, multiple profiles of IEEE 1588 have been developed by Standards Development Organizations (SDOs) to tailor the protocol to specific domains, for example, ITU-T's telecommunication frequency and phase profiles, the IEEE power-utility profile, and the SMPTE broadcast profile. As PTP deployments proliferate, the integrity of host-level timing has become critical, as minor errors can cascade into safety hazards in these high-stakes environments. 

Early versions of IEEE 1588 paid little attention to security, leaving the protocol vulnerable to various network-centric threats. Academic work soon demonstrated that an attacker who selectively delays or asymmetrically forwards \textsc{Sync}/\textsc{Follow\_Up} packets can induce clock offsets of hundreds of microseconds while remaining syntactically invisible to the slave servo \cite{finkenzeller2022feasible}. Other work showed that adversaries can exploit the \texttt{Best Master Clock Algorithm} (BMCA) to win the election process and broadcast a forged time base or inject bogus \textsc{Correction Field} values without ever violating the PTP state machine \cite{finkenzeller2024ptpsec}. These findings drove the standards community to retrofit protection mechanisms, notably the message authentication extension of Annex K in IEEE 1588-2008 and the more comprehensive security framework of Annex P of the current IEEE 1588-2019 edition. Researchers have complemented these with external cryptographic wrappers (e.g., IPsec and MACsec) and with key-management schemes such as \texttt{NTS4PTP} \cite{langer2022nts4ptp}. Delay-anomaly detectors that cross-check redundant paths \cite{finkenzeller2024ptpsec} and in-switch monitoring of timestamp variance \cite{itkin2020} further raise the bar for external adversaries who tamper with PTP traffic on the wire. Yet, as we show next, these defenses offer no protection once an adversary gains privileged access to the host itself. 

We identify a crucial blind spot: attacks originating within the host that executes PTP. Virtually all existing countermeasures assume an uncompromised time-keeping subsystem. We  argue that an attacker can achieve kernel-level privileges via a privilege-escalation or code-injection vulnerability, and can manipulate \texttt{clock\_gettime()}, intercept hardware timestamps, or inject skew directly into the PTP servo without altering any network traffic. Cryptographic safeguards become ineffective once the adversary operates within the trusted computing base. Prior research has demonstrated the feasibility and impact of attacks using kernel vulnerabilities on critical systems \cite{kim2025cr}. Yet, apart from preliminary analyses, kernel-rooted time attacks remain underexplored in the PTP literature.  

\noindent \textbf{Our Contributions.}
This paper closes that gap by providing the first systematic study of kernel-rooted time attacks in IEEE 1588 hosts. 

\begin{enumerate}[noitemsep, topsep=0pt]
    \item \textbf{Attack model and implementation.}
    We catalogue common Linux vulnerabilities that enable privilege escalation and code injection, distilling them into three attack primitives: constant offset, progressive skew, and random jitter. We implement each as an in-kernel payload that tampers with time-keeping while leaving PTP traffic untouched. 
    \item \textbf{Impact analysis on PTP servos.}
    Using a controlled test-bed, we measure how these attacks disturb stock implementations of \texttt{ptp4l} and \texttt{phc2sys} servos.
\end{enumerate}

By exposing a new class of host-local threats and showcasing their impact, our findings motivate further research in hardening the PTP security and treating the kernel as a first-class element in the trusted timing chain. 
\section{Background} \label{sec:background}

The Linux PTP implementation is divided between the kernel and the user space. Modern network interface cards (NICs) typically include a dedicated Precision Hardware Clock (PHC). Linux exposes these clocks via character devices (e.g., \texttt{/dev/ptp0}), allowing user programs to perform clock operations (e.g., \texttt{clock\_gettime(), clock\_adjtime()}, using standard POSIX calls. Linux also supports hardware-level timestamping on network packets via the \ttt{SO\_TIMESTAMPING} API, reducing timestamp uncertainty and facilitating accurate synchronization with external time sources. 

\begin{figure}
    \centering
    \includegraphics[width=0.65\columnwidth]{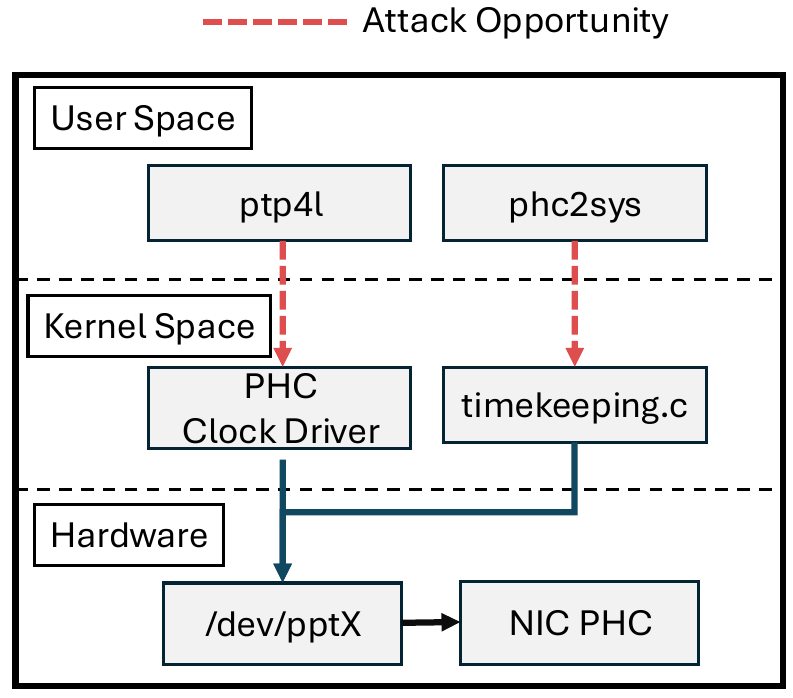}
    \caption{Architecture of the Linux PTP Stack}
    \vspace{-0.5cm}
    \label{fig:linux-ptp-stack}
\end{figure}

\subsection{Linux PTP Software Stack}

Figure \ref{fig:linux-ptp-stack} illustrates the Linux PTP software stack. The \ttt{linuxptp} suite, including \ttt{ptp4l} and \ttt{phc2sys}, leverages kernel-provided PHC and timestamping capabilities. \ttt{ptp4l} acts as an Ordinary or Boundary Clock per PTPv2 standard, synchronizing the NIC's PHC with a network master using hardware timestamps. Corrections are applied via \ttt{clock\_adjtime} kernel call. For a system lacking hardware support, \ttt{ptp4l} directly adjusts the system clock. The companion program \ttt{phc2sys} maintains the system clock aligned to the PHC, ensuring accurate system-wide time. 

\subsection{Kernel Timekeeping and Vulnerability Surface}

Linux kernel timekeeping maintains system clocks (e.g., \ttt{CLOCK\_REALTIME}, \ttt{CLOCK\_MONOTONIC}), using hardware counters and periodic interrupts, exposed via standard syscalls (\ttt{clock\_gettime(), clock\_settime(), adjtimex()}). Functions like \ttt{do\_settimeofday()} and \ttt{update\_wall\_time()} apply kernel-level time adjustments, critical for PTP synchronization. 


Common Linux kernel vulnerability classes have also been identified in its timing subsystems, including use-after-free bugs, missing pointer validations, and insufficient permission checks on ioctl interfaces. For instance, a use-after-free bug in a Linux timer handler arising from a race condition in \ttt{net/rose/rose\_timer.c} had allowed the attackers to crash the kernel and potentially execute code in kernel context \cite{CVE-2022-2318}. In the PTP clock driver itself, an unchecked function pointer was the root cause of a recent bug (CVE-2025-21814) where the PTP sysfs ioctl handles would call an uninitialized \ttt{.enable} callback, leading to a NULL pointer dereference. Although that particular flaw mainly causes a denial of service, it exemplifies how unguided ioctl paths in the PTP subsystem could be misused. Another example is an information leak in the \ttt{adjtimex()} API (CVE-2018-11508), where the compat ioctl (\ttt{compat\_get\_timex}) on 32-bit systems failed to initialize structure memory, allowing leakage of kernel data to user space. While not directly letting an attacker set the time, such a bug could be leveraged to glean information for a larger exploit that eventually gains kernel control. 


\section{PTP Kernel Vulnerabilities}\label{sec:threat-model}
Once attackers obtain kernel privileges, they can directly manipulate system or PHC clocks, bypassing standard capability controls\footnote{The Linux capability model requires CAP\_SYS\_TIME for time adjustments, but a kernel-mode payload can bypass these restrictions entirely.}. Such manipulation enables attackers to inject subtle offsets, skew clocks, or cause severe disruptions, undermining PTP servo stability and synchronization integrity. 

The critical points where an attacker can inject timing faults are the system call invocations used by the PTP daemons \ttt{ptp4l} and \ttt{phc2sys} to discipline time. The LinuxPTP implementation opens the PTP hardware clock device (e.g., \ttt{/dev/ptp0}) and treats it as a POSIX clock. An open file descriptor on a PHC device can be converted to a clock ID for use with \ttt{clock\_gettime}, \ttt{clock\_settime}, and \ttt{clock\_adjtime}. For example, during initialization \ttt{ptp4l} calls \ttt{phc\_open()} to open the NIC's PHC device and obtain a \ttt{clockid\_t} via the \ttt{FD\_TO\_CLOCKID} macro. Immediately after, \ttt{ptp4l} verifies the PHC is accessible by reading its current time with \ttt{clock\_gettime()} and ensures it can be adjusted by making a no-op adjustment call using \ttt{clock\_adjtime()}. If either fails, \ttt{ptp4l} aborts, as shown in Listing~\ref{lst:phc_open}.

\begin{listing}[H]
  \caption{Excerpt from \texttt{ptp4l} \texttt{phc\_open()}}
  \label{lst:phc_open}
  \begin{minted}[fontsize=\footnotesize]{c}
if (clock_gettime(clkid, &ts)) {
    close(fd);
    return CLOCK_INVALID;
}
if (clock_adjtime(clkid, &tx)) {
    close(fd);
    return CLOCK_INVALID;
}
  \end{minted}
\end{listing}

Once running, \ttt{ptp4l} (in client mode) uses hardware timestamping to measure the offset between the PHC and the server clock. The servo algorithm in \ttt{ptp4l} computes the correction and then directly adjusts the PHC using \ttt{clock\_adjtime()} on the PHC's clock ID. For small corrections, \ttt{ptp4l} slews the hardware clock by tuning its frequency: it populates a \ttt{timex} structure with \ttt{tx.modes = ADJ\_FREQUENCY} and a calculated \ttt{tx.freq} (in scaled ppm) before calling \ttt{clock\_adjtime}. This applies a subtle frequency correction (parts-per-billion level) to gradually pull the PHC in sync. For larger discrepancies, \ttt{ptp4l} performs a step adjustment: it uses the \ttt{ADJ\_SETOFFSET} mode to ask the kernel to atomically shift the PHC time by a specified offset. In the LinuxPTP code, this is done by setting \ttt{tx.modes = ADJ\_SETOFFSET | ADJ\_NANO} and filling \ttt{tx.time} with the offset to add (in seconds and nanoseconds), then calling \ttt{clock\_adjtime} on the PHC clock ID, as illustrated in Listing~\ref{lst:linuxptp_servo_step}.

\begin{listing}[H]
  \caption{Excerpt from LinuxPTP servo: prepare to step PHC clock by ns offset}
  \label{lst:linuxptp_servo_step}
  \begin{minted}[fontsize=\footnotesize]{c}
tx.modes = ADJ_SETOFFSET | ADJ_NANO;
tx.time.tv_sec  = sign * (ns / NS_PER_SEC);
tx.time.tv_usec = sign * (ns % NS_PER_SEC);
if (tx.time.tv_usec < 0) {
    tx.time.tv_sec  -= 1;
    tx.time.tv_usec += 1000000000;
}
if (clock_adjtime(clkid, &tx) < 0)
    pr_err("failed to step clock: %m");
  \end{minted}
\end{listing}

Using this offset, \ttt{phc2sys} adjusts the target clock. By default (when syncing the system to PHC). \ttt{phc2sys} adjusts \ttt{CLOCK\_REALTIME}. Like \ttt{ptp4l}, it will slew small offsets and step large ones. A proportional-integral (PI) servo in \ttt{phc2sys} computes a frequency correction in parts-per-billion (ppb) to apply to the system clock. \ttt{phc2sys} then invokes \ttt{clock\_adjtime(CLOCK\_REALTIME, \&tx)} with \ttt{tx.modes = ADJ\_FREQUENCY} to slew the system time. If the offset is beyond a step threshold (and stepping is permitted), \ttt{phc2sys} uses \ttt{ADJ\_SETOFFSET} to step the system clock similarly to the PHC case. In summary, \ttt{ptp4l} manipulates the NIC's PHC, while \ttt{phc2sys} manipulates the host's system time, both through the same kernel interfaces (just with different clock IDs and modes).

\begin{listing}[H]
  \caption{Compute PHC-to-system clock offset}
  \label{lst:phc_sys_offset}
  \begin{minted}[fontsize=\footnotesize]{c}
struct timespec tsrc, tdst;
if (clock_gettime(clkid, &tsrc))
    perror("clock_gettime");
if (clock_gettime(sysclk, &tdst))
    perror("clock_gettime");
offset = tdst.tv_sec * NS_PER_SEC - 
         tsrc.tv_sec * NS_PER_SEC +
         tdst.tv_nsec - tsrc.tv_nsec - rdelay;
  \end{minted}
\end{listing}

\subsection{Attack Surface}
The key points where an attacker can inject timing faults are the system call invocations for reading or adjusting time. As shown above, the \ttt{ptp4l} and \ttt{phc2sys} frequently call \ttt{clock\_gettime()} (to read current time) and \ttt{clock\_adjtime()} (to discipline clocks). These calls form the control flow junctures that directly influence the clock state. A kernel-rooted adversary, one who has gained privileged code execution in the OS, can exploit these junctures to poison the synchronization process. Specifically, an attacker can target:

\noindent $\bullet$ \textbf{Time Reads}: Every time the daemons call \ttt{clock\_gettime} on a clock, they trust the returned timestamp. Intercepting this allows an attacker to feed false time measurements into the servo loop. 

\noindent $\bullet$ \textbf{Time Adjustments}: Calls to \ttt{clock\_adjtime} carry the daemon's intended corrections. Hijacking these calls means the attacker can override the correction applied to the clock.

Because LinuxPTP uses the kernel's POSIX clock interface for PHC and system time, these become choke points that attackers can hook. The code excerpts in Listings~\ref{lst:phc_open}, \ref{lst:linuxptp_servo_step}, and \ref{lst:phc_sys_offset} highlight where the daemon hands off control to the kernel to apply a time correction. If an adversary can intercede at those points, they effectively control how time is adjusted.

\subsection{Vulnerability types}
\noindent $\bullet$ \textbf{Code Injection} vulnerabilities allow attackers to execute arbitrary code within the kernel space, providing direct access to manipulate timekeeping mechanisms and bypass standard security protocols. Such vulnerabilities significantly elevate the risk of undetected time synchronization manipulation. 

\noindent $\bullet$ \textbf{Privilege Escalation} vulnerabilities enable attackers with lower privileges to acquire kernel-level access. Once inside the kernel, adversaries can interfere with critical timekeeping operations, impacting synchronization precision and potentially causing broader systemic failures. 

\subsection{Attack Strategies}
This section presents three attack strategies that an adversary using privilege escalation or code injection kernel vulnerabilities can deploy to subtly corrupt local timekeeping on a PTP client host. By operating entirely within the host, these methods bypass traditional PTP packet-layer security. Notably, the approaches are generic and decoupled from any specific CVE or bug, i.e., any mechanism that grants kernel privilege or allows an adversary to inject code would suffice to carry out these attacks. The attacks are categorized by their effect on the clock: a constant offset injection introduces a steady bias, an incremental skew induces slow, long-term drift, and randomized delays inject timing noise and instability. Each has a distinct design goal (constant bias, drift, or instability) and degrades the PTP servo's performance in different ways, all while evading detection by PTP's on-network defenses. 

\subsubsection{Constant Offset Injection}
The \textbf{constant offset injection} attack is designed to impose a fixed time bias on the local clock. The attacker's goal is to create a persistent time error (e.g., a constant $\Delta t$ offset) between the host and the true time, thereby causing the PTP synchronization to converge to a false baseline. Essentially, the client clock is tricked into running consistently ahead or behind by a constant amount. This attack mimics the effect of a static propagation delay asymmetry or calibration error, except that the offset is maliciously introduced at the host. The PTP servo may attempt to correct the small offsets over time. However, if the attacker continually re-injects the bias, the servo will never eliminate the error; instead, the client clock will stabilize with a steady timing bias. 

Using this strategy, the local system clock can be offset by a constant value at all times. The design goal is a stable bias that does not obviously fluctuate and thus appears as a normal (if slightly wrong) time. This degrades the PTP servo's performance by ensuring it can never zero out the measured offset. The servo's control loop will continuously see an error of approximately $\Delta t$ and either keep adjusting (if $\Delta t$ is too large) or, if the offset is small enough, possibly settle, thinking the residual error is due to uncorrectable link asymmetry. In either case, the clock is never truly synchronized, constantly offset by $\Delta t$. Because this fault is maintained locally, PTP packet-layer defenses are blind to it. All incoming \ttt{Sync} and \ttt{Follow\_Up} messages from the grandmasters are processed normally and carry valid timestamps; only the local clocks' value is skewed. Even authenticity or delay attack detectors cannot flag a constant bias introduced at the host, since no protocol violation occurs \cite{alghamdi_precision_2021}. 
The attack is thus quite subtle: the system appears normally locked to the server except for a fixed offset in time, which could quietly undermine time-critical applications relying on absolute accuracy, such as trading timestamps or sequence ordering. 

In our implementation, we intercept the \ttt{clock\_gettime()} calls and add a fixed offset to the returned time. When \ttt{ptp4l} reads the PHC time as part of timestamping or when \ttt{phc2sys} reads the system time, our hook returns $t + \Delta$ instead of the true time $t$. Subsequently, the daemon calculates that the clock is $\Delta\mu s$ ahead or behind. Consequently, the servo will drive the clock toward this incorrect time. The system clock will maintain roughly a $\Delta$ offset indefinitely. In another attack, we intercept the adjustment calls: when \ttt{ptp4l} attempts to step the PHC by some offset, we inject a modified offset in \ttt{timex}. This modified offset reaches the kernel when \ttt{ptp4l} issues a \ttt{ADJ\_SETOFFSET}. This one-time tampering introduces a constant error that \ttt{ptp4l} thinks is corrected. Because \ttt{ptp4l} will now see no offset from the true time, it will maintain that $\Delta$ error. The code snippet in \ref{lst:linuxptp_servo_step} shows where such an offset can be injected. The attack effectively locks in a constant time fault by targeting the code path where the daemon thinks it is ``zeroing out" an offset. 

\subsubsection{Progressive Clock Skew}
The \textbf{progressive skew attack} aims to gradually deviate the client clock's rate or offset over an extended period, producing a slow-burn loss of synchronization. Instead of a one-time jump, the attacker introduces a small time error that grows cumulatively, causing long-term drift. The design goal is to create a creeping divergence that can eventually become significant (e.g., milliseconds or more). At the same time, at each individual synchronization interval, the deviation is subtle enough to avoid triggering alarms. This strategy deliberately imitates normal clock drift or environmental effects so that the PTP servo continually chases the moving target without suspecting malicious interference. Over time, the victim clock may wander far off the server's time (or even lose lock entirely) due to this incremental bias. 

Throughout a progressive skew attack, the key is that the induced error grows slowly; on each PTP sync interval, the offset change is within normal bounds. The servo will dutifully apply corrections, perhaps assuming the persistent slight drift is due to a low-quality oscillator or temperature effects. However, because the attacker continuously injects drift, the client clock's error will keep accumulating beyond what a healthy servo would typically allow. Eventually, the clock may exceed the permitted synchronization tolerance, resulting in noticeable resynchronization or even a failover (the PTP daemon might declare loss of synchronization if the offset grows too large). By that point, however, the compromise could have achieved its goal (e.g., causing a scheduled action to trigger too early/late or corrupting time-sensitive logs). Crucially, detecting this attack at the network level is difficult. No single Sync message or Offset value is malicious; each change is slight and could be attributed to benign sources. PTP's packet timing limits and outlier filters are typically tuned to ignore small jitter and drift, so an incremental skew that stays under these thresholds will evade immediate detection. 
This makes the progressive skew attacks a particularly insidious design, as it erodes synchronization gradually under the appearance of normal clock behavior. 

We implement this by manipulating frequency adjustments. Both \ttt{ptp4l} and \ttt{phc2sys} regularly apply small frequency tweaks via \ttt{clock\_adjtime(..., ADJ\_FREQUENCY)} to correct drift. We hook to these calls and insert a bias in the frequency values. Whenever \ttt{phc2sys} computes a frequency correction to align the system clock to the PHC, we modify the \ttt{tx.freq} field in the \ttt{timex} structure, multiplying the adjustment by a small factor. A subtle increase in this value each cycle will cause the system clock to run slightly fast. Over time, these small discrepancies accumulate into a significant error. There are other surfaces for this attack, for example, an attack could hook to \ttt{clock\_adjtime} itself and modify the frequency adjustment values, or adjust the registers carrying \ttt{tx.freq}. 

\subsubsection{Randomized Time Disturbances}
The third attack strategy focuses on injecting \textbf{random timing disturbances} into the host clock, with the goal of disrupting the stability of the PTP servo. Instead of a fixed or steadily growing offset, the attacker introduces noise (unpredictable fluctuations in the local timekeeping) to confuse the servo's control loop. The design goal here is stability disruption: by making the slave clock's error signal erratic, the attack prevents the PTP servo from achieving a steady state. The client clock will exhibit increased jitter and wander, degrading the quality of synchronization even if the average time offset remains bounded. In practice, this could cause oscillatory correction, frequent small time adjustments, or even oscillations around the correct time without converging, all of which can severely impact applications that require low timing variance (for example, control systems or high-speed financial transactions expecting a smooth clock). 

The net effect is that the local clock experiences a form of timing noise. From the PTP daemon's perspective, the offset measurements and frequency corrections become noisy, and the servo's job of filtering and smoothing becomes harder. Typical PTP servos employ filters or damping to handle normal network jitter, but an attacker can push this to extremes. The clock might keep switching between slight lead and lag relative to the server, causing the servo to constantly adjust (in effect, over-correcting one moment and under-correcting the next). In control system terms, the loop is forced into a high-noise regime, which can increase the Time Interval Error and MITE (Maximum Time Interval Error) beyond acceptable limits. Importantly, this volatility is not easily flagged as an attack. Random disturbances resemble the kind of packet delay variation or oscillator noise that PTP systems are expected to tolerate, just at a maliciously amplified level. 

From a security standpoint, 
the attack hides in the ``noise floor" of the timing system. Only careful monitoring of clock stability (e.g., a clock health monitor noticing abnormally high jitter) would \textit{hint} at a problem, and many PTP installations focus on accuracy and not on detecting an intelligent adversary injecting noise. Thus, randomized time fault is a subtle design to sabotage clock stability without overtly tripping security mechanisms. Even though all PTP messages are authentic and unaltered, the client clock's behavior is erratic, a condition that conventional PTP security extensions do not readily address. 

Our implementation injects noise into the time readings on a sporadic schedule. We attach a probe on \ttt{clock\_gettime}, that every $N$ calls, adds a random number of $\Delta\mu s$ to the returned timestamp. \ttt{phc2sys} then observes a sudden jump or drop in the offset, and overcorrects. In the next cycle, no noise is injected, causing an opposite correction, effectively the system clock jitters around the true time. In another attack, we insert random faults by intercepting \ttt{ADJ\_FREQUENCY}. Since the PTP servos operate continuously, even these infrequent random errors cause the servo to constantly hunt, never reaching a stable lock. From the daemon's perspective, the offsets and delay measurements become noisy, and the resultant clock adjustments lead to time jitter that is difficult to filter out.

In summary, all three attack strategies exploit the lack of secure binding between PTP's network signals and the host's actual clock state, by maliciously misusing the host's privileged clock control interfaces (e.g., \ttt{clock\_settime}, \ttt{clock\_adjtime}, PHC ioctls), an attacker can poison the local time smoothly and covertly. Each strategy targets a different aspect of PTP servo's operation: a constant offset defeats offset correction convergence, a slow skew thwarts long-term accuracy, and random disturbances undermine stability. 

\section{Evaluation} \label{sec:eval}

Our evaluation testbed consisted of two Beaglebone Black devices - a low-cost development board with a NIC that supports hardware timestamping, running the standard Debian distribution of the Linux operating system. We run the standard LinuxPTP suite using \ttt{ptp4l} and \ttt{phc2sys}, and configure the devices so that one acts as a server, and one as the client clock. Throughout our experiments, both daemons log at $1$ Hz with the \ttt{-m} (monitor) flag, recording server time ($t_{server}$), client clock time ($t_{client}$), measured offset, servo corrections and actual offset per interval. We evaluated four distinct scenarios: a baseline scenario without any attack; constant offset injection with a fixed bias of \(\Delta = 3\,\mu\mathrm{s}\); a progressive skew injection at three incremental rates ($\kappa=0.01,0.05,0.01,\mu\mathrm{s/s}$); and random disturbances with noise \(\sigma = 0.5\,\mu\mathrm{s}\)\footnote{The full dataset, including CSV logs and parsing scripts, is publicly available at: https://anonymous.4open.science/r/ptp-security-7D07}.

\begin{figure}[t]
    \centering
    \includegraphics[width=0.75\columnwidth]{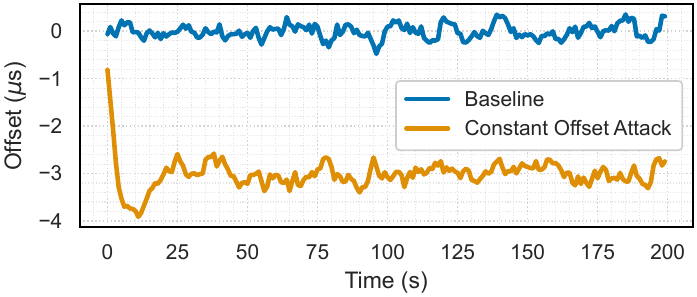}
    \vspace{-3mm}
    \caption{PTP Servo Offset under Constant Offset Attack.}
    \label{fig:ptp-constant-offset}
    \vspace{-0.4cm}
\end{figure}

Figure \ref{fig:ptp-constant-offset} plots the actual offset over time under a constant $3\mu s$ bias. The PTP servo converges quickly but settles with a steady residual offset of approximately \(\Delta_{\mathrm{res}} =\) \(3\)\,\(\mu\)s. The servo correction commands also stabilize near zero, indicating that the daemon believes it is fully synchronized despite the persistent error. 

\begin{figure}[t]
    \centering
    \includegraphics[width=0.75\columnwidth]{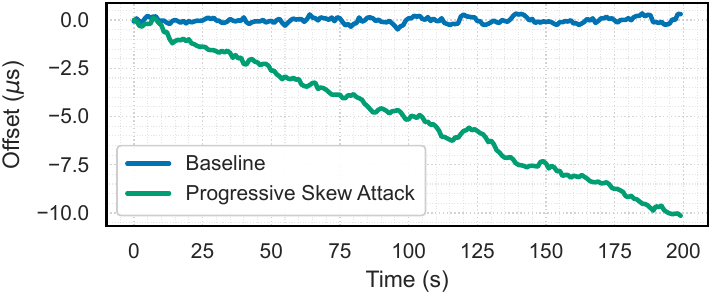}
    \vspace{-3mm}
    \caption{PTP Servo Offset under Progressive Skew Attack.}
    \label{fig:ptp-progressive-skew}
    \vspace{-0.5cm}
\end{figure}

Figure \ref{fig:ptp-progressive-skew} shows the client offset under an incremental skew of $0.05\mu s$. At $t=0$, the client is aligned; afterwards, the offset grows nearly linearly, reaching $10\mu s$ by $t=200$. The servo constantly applies minor corrections but cannot keep up with the drift. 
\begin{figure}[t]
    \centering
\includegraphics[width=0.75\columnwidth]{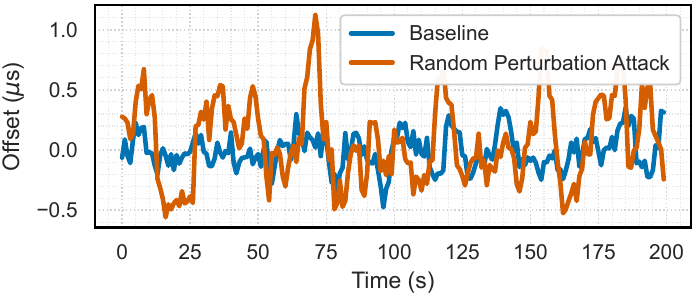}
\vspace{-3mm}
    \caption{PTP Servo Offset under Random Disturbance Attack.}
    \label{fig:ptp-random-perturbation}
\end{figure}
Figure \ref{fig:ptp-random-perturbation} shows the offset under random noise injections ($\sigma=0.5\mu s$). The client exhibits high-frequency jitter during the steady period ($t>50 s$). 
\begin{figure}[t]
    \centering
\includegraphics[width=0.75\columnwidth]{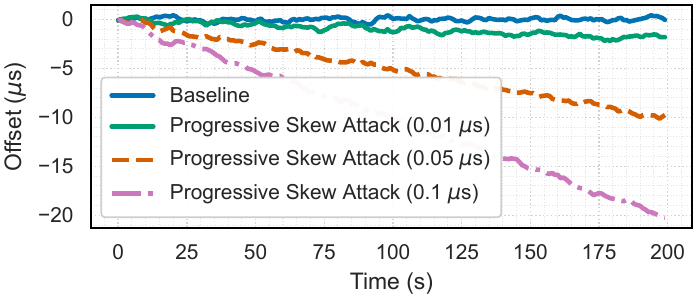}
\vspace{-3mm}
    \caption{PTP Servo Offset under Skew Attack with Multiple Skew Rates.}
    \label{fig:ptp-progressive-skew-multi}
\end{figure}
Figure~\ref{fig:ptp-progressive-skew-multi} overlays offsets for the skew rates of $0.01, 0.05, 0.1 \,\mu s/s$. The slope of each trace scales proportionally to the injection rate, demonstrating controllable degradation. 

Key takeaways from our experiments are that kernel-level manipulations can disrupt PTP synchronization without altering network packets. We observe that constant offsets lead to persistent errors that the servo falsely interprets as stable synchronization. Progressive skew attacks cause subtle, cumulative errors that significantly degrade synchronization. Random disturbances introduce high-frequency jitter, preventing stable servo convergence. 
Collectively, these results highlight critical vulnerabilities at the kernel syscall boundary, underscoring the ineffectiveness of network-layer defenses and emphasizing the need for securing host-level timekeeping.


\section{Related Work} \label{sec:related}

Previous security analyses of time synchronization have primarily focused on network-level vulnerabilities. For instance, several studies demonstrated attacks on the PTP where an adversary selectively delays or asymmetrically forwards \texttt{SYNC}/\texttt{FOLLOW\_UP} messages, silently inducing large clock offsets \cite{finkenzeller2022feasible}. Researchers have shown that attackers can exploit the Best Master Clock Algorithm (BMCA) in PTP by broadcasting falsified timestamps, severely skewing clocks while remaining compliant with protocol specifications \cite{finkenzeller2024ptpsec, alghamdi_precision_2021}.

Internal threats to time synchronization remain largely unexplored. Authors in \cite{alghamdi_precision_2021} categorize internal attackers with varying levels of privilege, emphasizing that advanced attackers could silently manipulate synchronization from within the host. Authors in \cite{moussa2019extension} presented detection and mitigation approaches for delay attacks targeting PTP synchronization in smart grid substation, highlighting that existing methods only partially cover the entire PTP attack surface. 

Recent literature has also investigated the impact of security mechanisms on PTP synchronization accuracy. For instance, integrating cryptographic mechanisms within PTP messages has been explored to ensure message integrity and authenticity \cite{fotouhi2023evaluation}. However, these mechanisms primarily defend against external threats and have limited effectiveness against internal attackers. Authors in \cite{fotouhi2023evaluation} experimentally evaluated these security controls on gPTP, revealing significant shortcomings against attackers who have gained internal network access.

More recent efforts have started exploring kernel-level attacks and trusted execution environments (TEEs) as a foundation for mitigating internal threats to timekeeping. For example, Authors in \cite{soomro2025time} demonstrated how Linux kernel interfaces can be exploited to corrupt timekeeping, highlighting the inadequacy of relying exclusively on external security measures. Meanwhile, Authors in \cite{nasrullah2024haest} proposed HAEST, which synchronizes clocks across heterogeneous IoT platforms by extracting timing cues from ambient events, removing the sole reliance on the kernel's clock services. In parallel, Timeguard \cite{nasrullah2024trusted} introduced a trusted time service within ARM TrustZone, providing secure and isolated access to hardware timers even in the presence of a privileged adversary. Together, these efforts emphasize the urgency of addressing internal kernel vulnerabilities and the promise of TEEs for building robust, end-to-end secure timing services. 

In summary, existing security efforts for PTP have mainly concentrated on external attackers and network-level defenses, with limited consideration for internal kernel-based threats. Our research expands the threat model by explicitly focusing on kernel-level vulnerabilities and internal attack strategies, emphasizing the need for deeper security integration spanning the entire synchronization stack, from network to kernel. 
\section{Conclusion} \label{sec:conclusion}

We presented a systematic study of kernel-level attacks on IEEE 1588 (PTP) synchronization. By exploiting kernel vulnerabilities, we demonstrated three in-kernel attack strategies that silently disrupt local synchronization without modifying network traffic. Our experimental results confirm that these stealthy manipulations significantly degrade synchronization accuracy, causing persistent errors, cumulative drift, or increased drift. This highlights a critical gap in current threat models: kernel-level adversaries can bypass existing cryptographic protections and anomaly detectors. Our findings underscore the need for integrated defenses addressing kernel-level threats. Operators should consider kernel compromise as seriously as traditional network threats in securing precision timing infrastructure.

\section*{Acknowledgment}
We thank the anonymous reviewers of IEEE ISPCS 2025 for their valuable feedback that helped improve this work. This work was supported by the National Science Foundation under Grant No. 2237485. 

\bibliographystyle{IEEEtran}
\bibliography{bibliography}

\end{document}